\pdfoutput=1
\documentclass[sigconf,table,xcdraw,english,bookmarks=false]{acmart}
\usepackage[utf8]{inputenc}

\usepackage{adjustbox}
\usepackage{csvsimple}
\usepackage{multirow}
\usepackage{amsmath}
\usepackage{longtable}
\usepackage[ruled,vlined,commentsnumbered,linesnumbered]{algorithm2e}
\usepackage{listings}
\usepackage{tcolorbox}
\usepackage{colortbl}
\usepackage{pgfplots}
\usepackage{pgfplotstable}
\def\code#1{\texttt{#1}}

\usepackage{subcaption}

\usepackage{xcolor}

\definecolor{codegreen}{rgb}{0,0.6,0}
\definecolor{codegray}{rgb}{0.5,0.5,0.5}
\definecolor{codepurple}{rgb}{0.58,0,0.82}
\definecolor{backcolour}{rgb}{0.95,0.95,0.92}
\definecolor{cellgreen}{HTML}{009901}
\definecolor{cellyellow}{HTML}{FFF700}
\definecolor{cellorange}{HTML}{FF6F00}
\definecolor{clgreen}{HTML}{34FF34}
\definecolor{clyellow}{HTML}{FDFF7E}
\definecolor{clorange}{HTML}{FE996B}
\definecolor{cellud}{HTML}{EE442F}
\definecolor{cellue}{HTML}{63ACBE}
\definecolor{celluo}{HTML}{CCBE9F}

\definecolor{cellgray}{HTML}{aeaeae}
 
\lstdefinestyle{mystyle}{
    backgroundcolor=\color{backcolour},   
    commentstyle=\color{codegreen},
    keywordstyle=\color{blue},
    numberstyle=\tiny\color{codegray},
    stringstyle=\color{codepurple},
    basicstyle=\ttfamily\tiny,
    breakatwhitespace=false,         
    breaklines=true,                 
    captionpos=b,                    
    keepspaces=true,                 
    numbers=left,                    
    numbersep=3pt,                  
    showspaces=false,                
    showstringspaces=false,
    showtabs=false,                  
    tabsize=2
}
 
\usepackage{array}
\newcolumntype{P}[1]{>{\centering\arraybackslash}p{#1}}

\copyrightyear{2020}
\acmYear{2020}
\setcopyright{acmlicensed}\acmConference[ASE '20]{35th IEEE/ACM International Conference on Automated Software Engineering}{September 21--25, 2020}{Virtual Event, Australia}
\acmBooktitle{35th IEEE/ACM International Conference on Automated Software Engineering (ASE '20), September 21--25, 2020, Virtual Event, Australia}
\acmPrice{15.00}
\acmDOI{10.1145/3324884.3415286}
\acmISBN{978-1-4503-6768-4/20/09}

\title{Using Defect Prediction to Improve the Bug Detection Capability of Search-Based Software Testing}

\author{Anjana Perera}
\email{Anjana.Perera@monash.edu}
\orcid{0000-0002-5080-9276}
\affiliation{%
  \institution{Faculty of Information Technology}
  \institution{Monash University}
  \city{Melbourne}
  \country{Australia}
}

\pgfplotsset{compat=1.16}

\begin{document}
\begin{NoHyper}

\begin{abstract}

Automated test generators, such as search based software testing (SBST) techniques, replace the tedious and expensive task of manually writing test cases. 
SBST techniques are effective at generating tests with high code coverage. However, is high code coverage sufficient to maximise the number of bugs found? 
We argue that SBST needs to be focused to search for test cases in defective areas rather in non-defective areas of the code in order to maximise the likelihood of discovering the bugs. Defect prediction algorithms give useful information about the bug-prone areas in software. Therefore, we formulate the objective of this thesis: \textit{Improve the bug detection capability of SBST by incorporating defect prediction information}. To achieve this, we devise two research objectives, i.e., 1) Develop a novel approach (SBST$_{CL}$) that allocates time budget to classes based on the likelihood of classes being defective, and 2) Develop a novel strategy (SBST$_{ML}$) to guide the underlying search algorithm (i.e., genetic algorithm) towards the defective areas in a class. Through empirical evaluation on 434 real reported bugs in the Defects4J dataset, we demonstrate that our novel approach, SBST$_{CL}$, is significantly more efficient than the state of the art SBST when they are given a tight time budget in a resource constrained environment.

\end{abstract}

\maketitle

\section{Introduction}

Search based software testing (SBST) has been gaining maturity not only in research, but also in the industry \citep{harman2015achievements,harman2018start}. Harman and Jones~\cite{harman2001search} coined the emerging research area, search based software engineering (SBSE), which uses search algorithms like genetic algorithm (GA) to solve software engineering problems. SBST is a sub-area of SBSE, which specialises in solving software testing problems like test data generation. Research on SBST dates back to 1976 \citep{miller1976automatic}, and since then it has been seen a growing interest by the research community \citep{harman2015achievements}. At the same time, SBST has made its way to the industry as well, i.e., Sapienz at Facebook \citep{alshahwan2018deploying,mao2016sapienz}.

Previous work shows that SBST techniques are effective at achieving high code coverage \citep{panichella2017automated, panichella2015reformulating}. However, the more important question is `How does SBST perform in finding real bugs?'. Both Shamshiri et al.~\cite{shamshiri2015automatically} and Almasi et al.~\cite{almasi2017industrial} showed that SBST outperformed other approaches in finding real bugs on open source and industrial projects respectively. However, overall the results of these studies suggest that SBST methods are not as effective in finding real bugs. We find that high code coverage is not sufficient to maximise the number of bugs found.

Defect prediction algorithms predict the areas in software that are likely to be buggy. They give predictions at file \citep{lewis2013does, de2015software, dam2019lessons}, class \citep{basili1996validation} or method \citep{caglayan2015merits, giger2012method, hata2012bug} level. Previous work shows the effectiveness of defect predictors in locating the buggy areas in software \citep{caglayan2015merits, nagappan2008influence, paterson2019empirical}. An effective defect predictor can be used to guide developers to efficiently test a software system to find bugs \citep{dam2019lessons}.

The objective of this thesis is to improve the bug detection performance of SBST by incorporating defect prediction information. We argue that SBST needs to generate more test cases covering the defective areas in the code in order to increase the likelihood of finding bugs. Therefore, we propose to leverage defect prediction information to inform the search process of the defective areas. To accomplish our main research objective, we formulate two research objectives; 1) Develop a novel approach (SBST$_{CL}$) that allocates time budget to classes based on the likelihood of classes being defective, and 2) Develop a novel strategy (SBST$_{ML}$) to guide the underlying search algorithm (i.e., genetic algorithm) towards the defective areas in a class. Finally, the expected contributions of this research are 1) a time budget allocation approach that improves bug detection capability of SBST, 2) an analysis of the bug detection performance improvement of SBST by the time budget allocation approach when varying the performance of the defect predictor, 3) a defectiveness-aware search algorithm guidance strategy to improve the bug detection capability of SBST, and 4) an analysis of the bug detection performance improvement of SBST by the defectiveness-aware search algorithm guidance strategy when varying the performance of the defect predictor.

Through empirical evaluation on 434 real bugs from 6 open source java projects in the Defects4J dataset~\citep{just2014defects4j}, we demonstrate the improved efficiency of SBST by our time budget allocation approach in a resource constrained environment. 
Further analysis to the results suggests the improvement of SBST$_{CL}$ comes from its ability to find more unique bugs which cannot be found otherwise.

\section{Motivating Example}

The following example illustrates the limitation of SBST focusing only on high code coverage when using it for finding real bugs. Figure~\ref{listing:math-94-buggy-code} shows a classic example of a bug due to an integer overflow. The \code{if} condition at line 412 is expected to evaluate to \code{true} only if either \code{u} or \code{v} is zero. However, if the method is called with the following inputs u = 1073741824 and v = 1032, then the \code{if} condition at line 412 is evaluated to \code{true} since the multiplication of \code{u} and \code{v} causes an integer overflow to zero. Fixed code shown in Figure~\ref{listing:math-94-fixed-code} rectifies this issue by individually checking if \code{u} or \code{v} are zero. Thus, in order to detect this bug, a test should pass non-zero arguments \code{u} and \code{v} such that their multiplication causes an integer overflow to zero. A search algorithm~\cite{panichella2017automated} focusing on high code coverage is more likely to find a test case with both/either u and/or v = 0 and consider this code as covered than find a test case with a pair of u and v causing an integer overflow to zero. In contrast, we propose to 1) use defect prediction information to identify bug-prone areas in the code and 2) guide the search algorithm to find more test cases covering the bug-prone code to increase the likelihood of finding the bug.

\begin{figure}[h]
\vspace{-2mm}
\begin{subfigure}{0.20\textwidth}
\begin{lstlisting}[language=java, firstline=411, lastline=416, firstnumber=411, xleftmargin=0em, columns=fullflexible]
public static int gcd(int u, int v) {
	if (u * v == 0) {
		return (Math.abs(u) + Math.abs(v));
	}
	...
}
\end{lstlisting}
\caption{Buggy code}
\label{listing:math-94-buggy-code}
\end{subfigure}\hspace{0.03\textwidth}
\begin{subfigure}{0.20\textwidth}
\begin{lstlisting}[language=java, firstline=411, lastline=416, firstnumber=411, xrightmargin=-2em, columns=fullflexible]
public static int gcd(int u, int v) {
	if ((u == 0) || (v == 0)) {
		return (Math.abs(u) + Math.abs(v));
	}
	...
}
\end{lstlisting}
\caption{Fixed code}
\label{listing:math-94-fixed-code}
\end{subfigure}
\caption{MathUtils class from Math-94 Bug in Defects4J}
\vspace{-4mm}
\end{figure}

\section{Related Work}

Shamshiri et al.~\cite{shamshiri2015automatically} and Almasi et al.~\cite{almasi2017industrial} evaluated EvoSuite~\citep{fraser2012whole}, a state of the art SBST tool, against other test generation techniques on open source and industrial software projects respectively. While the results of these studies showed that EvoSuite outperformed the other techniques, overall EvoSuite is not as effective in finding real bugs. In particular, EvoSuite only found on average 23\% bugs from the Defects4J dataset \citep{shamshiri2015automatically}. Both studies used only 100\% branch coverage criterion for the test case search. We argue that targeting 100\% code coverage is not sufficient to find real bugs. Instead, we propose to guide SBST to search for test cases covering defective areas in code to maximise the bug detection. Contrary to these works, we leverage defect prediction information to guide the search algorithm to extensively explore the search space for test cases covering defective areas.

Defect predictors employ statistical approaches~\cite{caglayan2015merits, nagappan2008influence}, learning techniques~\cite{hall2011systematic, giger2012method} or risk estimation algorithms~\cite{lewis2013does, paterson2019empirical, googledefect} etc. to estimate the likelihood of a file, class or method is defective. Based on this likelihood and a threshold (e.g., 0.5), they classify if the component (i.e., class, file or method) is defective or not. As a result of their efficacy, defect predictors are used in the industry to assist developers in code reviewing \cite{googledefect} and guide the developers to efficiently test defective components \citep{dam2019lessons}. Defect predictors can be useful not only for humans, but also for other automated tools. Paterson et al.~\cite{paterson2019empirical} used a defect predictor to inform a test case prioritisation strategy, G-clef, of the classes that are likely to be defective and found it is promising. While the predictions are beneficial for developers, the inaccuracies in the estimations (e.g., false positives) can become costly and inefficient. Our approaches take this into account when using defect predictions to inform SBST.

\section{Research Objectives}

\vspace{-1mm}
\begin{tcolorbox}
RO1: Develop a novel approach (SBST$_{CL}$) that allocates time budget to classes based on the likelihood of classes being defective
\end{tcolorbox}
\vspace{-1mm}

SBST techniques such as the ones implemented in EvoSuite~\citep{fraser2012whole} employ a search algorithm such as genetic algorithm (GA)~\citep{fraser2011evolutionary} to generate unit tests for each class independently by spending a given time budget. This time budget has to be tuned carefully as it is a stopping criterion for the GA and depends on the available computational resources in the organisation. Allocating a higher time budget is beneficial for the search because it allows the search algorithm to extensively explore the search space of test cases, thereby increasing the likelihood of finding bugs.

Previous work that studied SBST in terms of finding real bugs allocated a fixed time budget to all the buggy classes \citep{almasi2017industrial, shamshiri2015automatically, gay2017generating}. Ideally, buggy classes should receive more time budget than the non-buggy ones to maximise the likelihood of finding bugs. Allocating different time budgets to classes has been explored in previous research. For example, Campos et al.~\citep{campos2014continuous} allocated time budget to classes based on the complexity of them in order to maximise branch coverage. In contrast, we propose to allocate time budgets based on the likelihood of a class being buggy, i.e., classes that are more likely to be buggy receive a higher time budget, in order to increase the number of bugs found.

In the first research objective, we simulate defect predictor outputs at different performance levels (e.g., Matthews Correlation Coefficient (MCC)) including an ideal defect predictor to investigate how the bug detection performance of SBST$_{CL}$ changes with varying performance levels of defect predictors. An actual defect predictor comes with some degree of inaccuracy. Thus, we investigate the bug detection performance of SBST$_{CL}$ when using a real defect predictor at class level.

\begin{tcolorbox}
RO2: Develop a novel strategy (SBST$_{ML}$) to guide the underlying search algorithm (i.e., genetic algorithm) towards the defective areas in a class
\end{tcolorbox}

SBST techniques receive a limited time budget to generate unit tests for a class. Hence, spending search resources on covering non-defective parts in the class will deteriorate the bug detection performance of SBST. GA uses search heuristics (e.g., fitness function and selection operator) to guide the search to the given objective. Current state of the art search algorithm, $DynaMOSA$~\cite{panichella2017automated}, employs a preference sorting algorithm to sort the best test cases to be selected to the next generation in GA. We find that preference sorting treats each target (i.e., coverage goal) is equally important and that leads to sub-optimal test suites in terms of finding bugs. Thus, we plan to use defect prediction at method level to inform the underlying GA of the buggy methods, hence it can prioritise the coverage of buggy code over non-buggy code.

As in the RO1, we simulate defect predictor outputs to investigate how the bug detection performance of SBST$_{ML}$ changes when varying the performance of the defect predictor. 
Then, we investigate the performance of SBST$_{ML}$ when using an actual defect predictor at method level.

\section{Current State of Work}

\begin{figure}[!ht]
    \centering
    \includegraphics[width=0.48\textwidth]{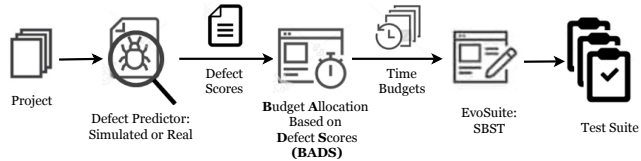}
    \caption{SBST$_{CL}$ Overview}
    \label{fig:prioritizer}
\end{figure}

\subsection{RO1: Develop a novel approach (SBST$_{CL}$) that allocates time budget to classes based on the likelihood of classes being defective}

\subsubsection{Motivation}

Real-world projects are often very large. For instance, a modern car has millions of lines of code and thousands of classes \citep{broy2007engineering}. Thus, running test generation on each individual class in the project would take significant amount of resources. At the same time, the available computational resources are limited in practice \citep{campos2014continuous}. In order for the test generation tool to be adapted in the continuous integration (CI) system \citep{fowler2006continuous} without causing an impact to the existing processes (e.g., making other jobs idle), the test generation tool should use as little resources possible. Therefore this prompts the question, `How can we optimally utilise the available computational resources (e.g., time budget) to generate tests for the whole project with maximal bug detection?'.

\subsubsection{Approach} 

We propose our novel approach, SBST$_{CL}$, consisting of three main modules (see Figure \ref{fig:prioritizer}); i) Defect Predictor, ii) Budget Allocation based on Defect Scores (BADS), and iii) SBST. The defect predictor estimates the likelihood of defectiveness for each class in the project. We call this likelihood a defect score. BADS takes these defect scores as input and decides how to allocate the available time budget to classes based on these scores. At high level, BADS allocates more time budget for more likely to be defective classes and less time budget for less likely to be defective classes. BADS employs an exponential function to highly favour the budget allocation for a few number of highly likely to be defective classes. In order to account for inaccuracies present in defect predictors, we introduce a lower bound for time budget allocation. This lower bound ensures that every class gets a time budget allocated regardless of its defect score. Finally, the SBST module takes these allocated time budgets and runs test generation for each of the classes. Given its maturity, we use EvoSuite as the SBST tool and $DynaMOSA$ as the search algorithm.

\subsubsection{Evaluation}
\label{subsec:RO1_evaluation}

We answer the following research questions in order to evaluate SBST$_{CL}$ in terms of efficiency in finding real bugs and effectiveness of discovering unique bugs, i.e., bugs that are not found by the baseline approach.

\begin{center}
    \textit{RQ1.1: Is SBST$_{CL}$ with simulated defect predictors more efficient in finding bugs compared to the state of the art?}
\end{center}

To investigate the bug detection performance of SBST$_{CL}$ when employing defect predictors with different performance levels, we simulate the defect predictor outputs. Studies on defect prediction use various metrics to evaluate and compare the models' performance, i.e., recall, precision, accuracy, MCC and AUC etc.~\cite{hall2011systematic, hosseini2017systematic}. We choose MCC as the performance measure to simulate the defect predictions as it is insensitive to class imbalance unlike the other metrics~\citep{shepperd2014researcher}. Class imbalance prevails in defect prediction as there are only a few number of buggy classes (or methods) compared to non-buggy ones. We simulate defect predictor outcomes ranging from MCC = 1.0 (i.e., ideal defect predictor) to MCC = 0.0 (i.e., random classification).

We choose Defects4J as a suitable benchmark to evaluate our approach as it is widely used in research~\citep{shamshiri2015automatically, le2016history, paterson2019empirical, pearson2017evaluating}, and hence allows us to compare the results to existing work. It contains 835 real bugs from 17 open source java projects~\citep{defects4jweb}. The baseline for comparison is the state of the art SBST, i.e., $DynaMOSA$, with fixed time budget allocation. We conduct experiments for 2 cases of total time budgets ($T$); $15*N$ and $30*N$ seconds, where $N$ is the number of classes in the project. We choose these time budgets for the experiments as $DynaMOSA$ is capable of converging to the final branch coverage quickly, sometimes with a time budget like 20 seconds \citep{panichella2017automated}. Also, faster test generation makes SBST more suitable to fit into the CI/CD pipeline without affecting the existing processes.

For each bug, we run test generation with the baseline and SBST$_{CL}$ using each simulated defect predictor. We repeat the test generation of each approach for 20 runs to account for randomness of SBST. Once the test cases are generated, we check if they find the bugs in the programs, and compare the results of SBST$_{CL}$ with different simulated predictors against the baseline. We use two-tailed non-parametric Mann-Whitney U-Test with the significance level ($\alpha$) 0.05 \citep{arcuri2014hitchhiker} and Vargha and Delaney's $\widehat{A}_{12}$ statistic \citep{vargha2000critique} to check the statistical significance of the differences and effect size of the results (i.e., number of bugs found).

\begin{center}
    \textit{RQ1.2: Does the efficiency of SBST$_{CL}$ change when using defect predictors with different performance levels?}
\end{center}

To answer this research question, we compare SBST$_{CL}$ with simulated defect predictors (MCC $<$ 1.0) against SBST$_{CL}$ with the ideal defect predictor (i.e., MCC = 1.0). SBST$_{CL}$ considers the inaccuracies present in the actual defect predictors when allocating time budgets to classes. However, the defect predictors having performance closer to a random classifier (i.e., MCC = 0.0) may deteriorate the performance of SBST$_{CL}$.

\begin{center}
    \textit{RQ1.3: Does SBST$_{CL}$ with simulated defect predictors find more unique bugs?}
\end{center}

In this research question, we investigate if SBST$_{CL}$ is capable of finding more unique bugs which cannot be revealed by the baseline approach. This is an important measure since it indicates the testing technique is capable to find bugs which cannot be discovered otherwise within the given time budget ~\citep{habib2018many}. The increased efficiency of SBST$_{CL}$ could also be due to its robustness, which is measured by the success rate. Thus, we also analyse how often a bug is found over 20 runs.

\begin{center}
    \textit{RQ1.4: Is SBST$_{CL}$ with a real defect predictor more efficient and effective in finding bugs when compared to the state of the art?}
\end{center}

The first three research questions (i.e., \textit{RQ1.1-3}) investigate the bug detection performance of SBST$_{CL}$ using simulated defect predictors. In this research question, we investigate the performance of SBST$_{CL}$ when using an actual defect predictor.

We use Schwa~\citep{de2015software, schwagithub}, which uses three metrics that have been shown to be effective in the literature~\citep{rahman2011bugcache, kim2007predicting, graves2000predicting, nagappan2008influence}, as the defect predictor. Schwa produces defect scores for each class in the project by using a time weighted risk formula, whereas the other approaches require to train a classifier using training data. This makes it ideal to use in an industrial scenario where such training data is not available. In addition, Paterson et al.~\cite{paterson2019empirical} successfully leveraged Schwa in G-clef to inform a test case prioritisation strategy of the buggy classes.

We run experiments for SBST$_{CL}$ with Schwa (SBST$_{CL+Schwa}$) as outlined in \textit{RQ1.1}. Then, we compare the efficiency and effectiveness (i.e., finding unique bugs) of SBST$_{CL+Schwa}$ against the state of the art SBST (i.e., baseline approach).

\subsubsection{Results} 

We are yet to complete the experiments for \textit{RQ1.1-3}. For \textit{RQ1.4}, the empirical evaluation demonstrates that SBST$_{CL+Schwa}$ is significantly more efficient than state of the art SBST when they are given a tight time budget ($T$ = $15*N$ seconds) in a usual resource constrained scenario. In particular, SBST$_{CL+Schwa}$ finds an average of 13.1\% more bugs than the baseline method. However, when there is sufficient time budget ($T$ = $30*N$ seconds), SBST$_{CL+Schwa}$ is more effective than the state of the art SBST 67\% of the time (effect size = 0.67), which is significant given how hard it is to find failing test cases \citep{habib2018many}. Further analysis to the results tells us that the superior performance of SBST$_{CL+Schwa}$ is supported by both its ability to find unique bugs that the baseline could not find and robustness in finding bugs. More details can be found in our publication~\citep{perera2020defect}.

\subsection{RO2: Develop a novel strategy (SBST$_{ML}$) to guide the underlying search algorithm (i.e., genetic algorithm) towards the defective areas in a class}

\subsubsection{Motivation}

We establish in RO1 that the limited resources (i.e., time budget) available in practice can be allocated optimally for test generation of classes in order to maximise the number of bugs found. However, does the search algorithm spend the allocated resources (i.e., time budget) to optimally search for test cases that can reveal the bugs? We find that the current state of the art search algorithm, $DynaMOSA$, treats each target (i.e., coverage goal) is equally important to cover. However, the buggy code could exist in one or few methods, hence we argue that spending the search resources on covering the non-buggy methods may lead to sub-optimal test suites in terms of finding bugs.

\subsubsection{Approach} 

We use a defect predictor that works at the method level. It estimates the likelihood of defectiveness (i.e., defect score) for each method in a class. We choose state of the art $DynaMOSA$ as the search algorithm in SBST$_{ML}$. First, SBST$_{ML}$ uses the defect scores to identify the buggy methods that need to be prioritised in the search. Then, all the targets (e.g., branches) in each buggy method are considered equally important. We introduce a defectiveness-aware preference sorting algorithm that prioritises the selection of test cases that are closer to cover the buggy targets to the next generation in GA. This way SBST$_{ML}$ can generate more test cases covering the buggy targets in the class, hence increasing the likelihood of finding the bug.

\subsubsection{Evaluation} 
\label{subsec:RO3_evaluation}

We mainly evaluate SBST$_{ML}$ in terms of efficiency in finding real bugs and effectiveness of revealing unique bugs against the state of the art, $DynaMOSA$. Similar to RO1, we plan to carry out the investigation in two parts. First, we study if the performance of SBST$_{ML}$ changes when using simulated defect predictors with varying performance levels. Second, we investigate the performance of SBST$_{ML}$ when using an actual defect predictor that works at method level. 

We use the Defects4J dataset as the experimental subjects. Similar to the experimental protocol outlined in RO1, we run test generation for each bug with the the baseline and SBST$_{ML}$ using simulated and actual defect predictors. We repeat the test generations for 20 runs and conduct statistical tests to check for significance of the differences of results (i.e., number of bugs found).

\subsubsection{Results}

We conducted a pilot study on 60 bugs from Apache-commons Lang project~\citep{defects4jweb}. Our preliminary results show that SBST$_{ML}$ with an ideal defect predictor is significantly more efficient than $DynaMOSA$. In particular, SBST$_{ML}$ finds an average of 11.13\%, 20.5\% and 21\% more bugs than $DynaMOSA$ at 15, 30 and 60 seconds respectively.

\section{Future Work}

We have three future works pertaining to RO1 (\textit{RQ1.1-3}) and RO2. In the first research objective, we need to perform the experiments for SBST$_{CL}$ with simulated defect predictors and evaluate its performance as outlined in Section \ref{subsec:RO1_evaluation}. 

Second and third future works aim at achieving the second research objective. The outcome of the pilot study suggests the merits of using defect prediction at method level to guide the search algorithm to find more bugs. Our next step is to incorporate the lessons learned from the pilot study. This is followed by an empirical evaluation as outlined in Section \ref{subsec:RO3_evaluation}. Upon completing the second future work, we apply an actual defect predictor that works at the method level. The existing work on method level defect prediction suggests several metrics that are effective in producing predictions \citep{giger2012method, hata2012bug, shippey2019automatically}. We plan to use these metrics to build a model and use it in the SBST$_{ML}$.

\section{Conclusion}

In this thesis, we aim at achieving the main research objective; \textit{Improve the bug detection capability of SBST by incorporating defect prediction information}. Through empirical evaluation on 434 real bugs in Defects4J dataset, we demonstrate the increased efficiency of SBST by our time budget allocation approach in a resource constrained environment. Preliminary results of the second research objective shed a positive light on guiding the search algorithm using the likelihoods of methods being defective in order to find more bugs. The outcome of \textit{RQ1.4} has been accepted at the International Conference on Automated Software Engineering (ASE)~\citep{perera2020defect}.

\clearpage
\bibliographystyle{ACM-Reference-Format}
\bibliography{reference}

\end{NoHyper}
\end{document}